\documentstyle[12pt,epsf]{article}
\begin{document}

\title{Nuclear Transport at Low Excitations}
\author{Helmut Hofmann
$^{1}$ and Fedor A.Ivanyuk$^{2}$\\ \small\it{1) Physik-Department
der TU M\"unchen, D-85747 Garching, Germany \thanks{
e-mail:hhofmann@physik.tu-muenchen.de} }
\\
\small\it{2) Institute for Nuclear Research, 252028 Kiev-28,
Ukraine \thanks{e-mail: ivanyuk@kinr.kiev.ua} }}
 \maketitle
\begin{abstract}
Numerical computations of transport coefficients at low
temperatures are presented for shapes typically encountered in
nuclear fission. The influence of quantum effects of the nucleonic
degrees of freedom is examined, with pair correlations included.
Consequences for global collective motion are studied for the case
of the decay rate. The range of temperatures is specified above
which this motion may be described as a quantal diffusion
process.\end{abstract}
\medskip

\centerline{PACS numbers: 05.60.Gg, 24.10.Pa, 24.75.+i, 25.70.Ji}
 \bigskip
\centerline{Phys. Rev. Lett. 82 (1999) 4603}
 \vspace{20pt}

In the past decade much progress has been made in the
understanding of nuclear transport phenomena in the regime of not
too low temperatures, say between 1 and 5 MeV (with $k_B=1$). Such
a situation is reached if two heavy ions collide at an energy
above the Coulomb barrier, but where the excess energy per
particle still is small compared to the Fermi energy.  In this
regime the dynamics of the composite system may be parameterized
in terms of shape variables. Of particular interest is the
outgoing channel which is dominated by fission and the emission of
light particles and $\gamma$'s. It has been possible
experimentally to deduce valid information on the time scale of
collective motion \cite{exptisc}, and, hence, on the size of
nuclear dissipation. These experiments suggest collective motion
to be over-damped, possibly providing an answer to the question
raised by Kramers as early as 1940 in his seminal paper
\cite{kram}, namely whether nuclear friction is "abnormally small
or abnormally large". Nowadays such processes are described
theoretically in terms of the Langevin equation \cite{mactheo},
which is understood to be equivalent to Kramers' original equation
(of Fokker Planck type) for the density in collective phase space.

On general grounds, it may be anticipated that the magnitude of
nuclear dissipation will vary with excitation. Indeed, there are
experimental indications \cite{exptdepf} for such a conjecture. At
small thermal excitations the dynamics is governed by the (real)
mean field for which there is no room for damping of slow
collective motion like fission. At larger T coupling to more
complicated configurations sets in, which causes transfer of
energy from the collective degrees of freedom $Q_\mu$ to the
nucleonic ones $x_i$.  Within the linear response approach
\cite{hofrep} the effects of this coupling are accounted for by
dressing the single particle energies by complex self-energies
depending both on frequency and T. Approximating its imaginary
part by a constant proportional to $T^2$, friction will again
decrease with $T$, once the microscopic damping becomes so large
that one may speak of "collision dominance". At intermediate
temperatures there might be the intricate contribution to friction
from the "heat pole", which has been seen to be large for
non-ergodic systems \cite{hoivyanp}, and which has a dramatic
influence on the T-dependence.

In the present Letter we want to focus on very low excitations,
say in the range below $T\approx  1 {\;{\rm MeV}}$.  This regime
not only is of great practical importance, as for the production
of super-heavy elements \cite{supheav}, for instance, it is also
of great theoretical interest.  First of all, there is little
doubt that in this domain quantum effects dominate nucleonic
dynamics, and the transport coefficients will strongly be
influenced by shell effects and pair correlations.  One even must
expect quantum features to be present for collective motion, for
instance as corrections to Kramers formula for the decay rate
\cite{hofthoming}.  Often quantal approaches are based on the
functional integral method applied to simplified Hamiltonians of
the Caldeira-Leggett type (for a review see \cite{hantalbor}).
There, the bath degrees of freedom $x_i$ are represented by a set
of oscillators of fixed frequencies, with a bilinear coupling
between the $x_i$ and the collective variable $Q$. The decay rate
is calculated for imaginary time propagation. Both features hardly
can be taken over to nuclear fission. First of all, the simplest
constraint to warrant self-consistency between the mean field and
the shape of the nuclear density requires the former to change
with $Q$. This aspect alone makes it very difficult to work with a
(pre-fixed) Hamiltonian for the total system of all degrees of
freedom.  Moreover, the temperature, which one may define
\cite{temflu} for the fast degrees of freedom (supposedly be given
by the "nucleonic" ones), is subject to changes with $Q$ as well
as with time. The latter feature occurs because of the evaporation
of particles mentioned before.

For these reasons a formulation with real time propagation is
much more appropriate. This has been achieved by a suitable
application of linear response theory on the basis of a locally
harmonic approximation (LHA) (for a review see\cite{hofrep}).
One exploits the concept of propagators which move the system
forward in collective phase space by small time steps. As
the individual ones only cover small areas they may be
represented by (multi-dimensional) Gaussians. The latter satisfy
an equation of motion whose structure is similar to that of
Kramers, with only the diffusive terms being modified to account
for quantum effects.

The following study is based on numerical calculations of transport
coefficients for average motion, namely friction $\gamma$,
inertia $M$ and local stiffness $C$, more precisely of those
ratios which determine transport in phase space,
\begin{equation}\label{defgameta}
\Gamma_{\gamma}={\gamma \over M},\quad \varpi ^2=
   {{\mid C \mid }\over M},\quad \eta ={\Gamma_{\gamma} \over 2\varpi}
    = {\gamma \over 2\sqrt{M\mid C \mid}}
\end{equation}
Their knowledge will allow us to examine implications on the
diffusion coefficients and, hence, on transport processes like
fission.  It would be most desirable that such information be used
as input for computational codes to solve for Fokker-Planck or
Langevin equations. In this way one would be able to examine in
more detail the role of shell effects, which are known to produce
structure not only in the static energy but in the inertial and
frictional forces as well. To simplify matters, for the present
purpose we will look at the more schematic case where the system's
energy exhibits just one minimum and one barrier at $Q_a$ and
$Q_b$, respectively. The stiffnesses and the barrier height are
found from a Strutinsky calculation of the free energy. Finer
details of shell effects are removed both from the potential as
well as from the transport coefficients by applying a suitable
smoothing over a small region of the collective variable around
$Q_a$ and $Q_b$.

Suppose we may at first discard any quantum effects in the
collective degrees of freedom, which amounts to look at the "high
temperature limit", for which Kramers' equation applies
\cite{hofrep}. The temperatures we have in mind are always small
compared to the barrier height, $T\ll E_b$. The decay rate $R_K$
then shows the following behavior.  For given $T$, but as function
of the $\eta_b$ (at the barrier), the $R_K(\eta_b,T)$ increases
first, to decrease after it has reached a maximal value (see
e.g.\cite{scheuho}). The decreasing branch is represented well by
Kramers' "high viscosity limit" \cite{kram}
\begin{equation}\label{krahivi}
R_K={\omega_a\over 2\pi}
\left(\sqrt{1+\eta_b^2} - \eta_b\right) \exp(-E_b/T)
\end{equation}
which is valid for $\eta_b \geq {T/(2E_b)}$ (see
e.g.\cite{hantalbor}). If blindly extended down to $\eta_b=0$ this
form $R_K$ reaches a value typical for a simple transition state
model, $r_{TST}\approx r(\eta_b=0)$ (Bohr-Wheeler formula).
Rather, for very small $\eta_b$, one ought to apply Kramers' "low
viscosity limit", given by $R_K^{l.v.}=\Gamma_\gamma^b (E_b/T)
\exp(-E_b/T)$. For nuclear physics the latter has not played any
role yet, as $\eta$ is believed to lie above the limit given below
(\ref{krahivi}). According to  \cite{hofrep, yaivho, ivhopaya}
this should be the case at temperatures above $1 {\;{\rm MeV}}$.
Moreover, the $\varpi$ does not change much with $T$. In
\cite{ivhopaya} a value of about $1{\;{\rm MeV}}/\hbar$ was found
both at the potential minimum as well as at the saddle. It so
turns out that this feature is more or less recovered even at
smaller $T$, say within an accuracy of the order of $20 \%$, which
may be good enough for the following discussion.

More drastic modifications are expected, and indeed seen, for
dissipation. To study this behavior, the $\gamma, \Gamma_\gamma$
and $\eta$ have been  calculated on the basis of the same deformed
shell model as in \cite{ivhopaya} but with pair correlations
included. The transformation from independent particles to
quasi-particles of BCS-type is standard.  For our purpose the
common procedure does not suffice, however. As mentioned
previously, a decent and sensible description of nuclear
dissipation needs to account for "collisions".  At low thermal
excitations their effects, too, will strongly be influenced by
pair correlations. Look at the extreme case of zero temperature
and take an even-even system for the sake of simplicity. Then
there will be no quasi-particle states within twice the gap energy
$2 \Delta$. Hence, the imaginary part of the self-energy
$\Gamma(\hbar\omega, \Delta, T=0)$ must be zero at least within
such a range of frequencies $\hbar\omega$. Hence, at $T=0$
friction will strictly vanish for any collective motion whose
frequency  $\omega = \varpi$ lies in that range, i.e. $\hbar
\omega \leq 2\Delta$.  Extrapolating from the case of $T=0$ to
finite $T$ we should expect the function $\gamma=\gamma(T)$ to
have a step like behavior. This dependence then goes through to
that of $\Gamma_\gamma$ and $\eta$, albeit the inertia, too, is
influenced by pairing.

\begin{figure}[htb]
\centerline{{
\epsfxsize=8.3cm
\epsffile{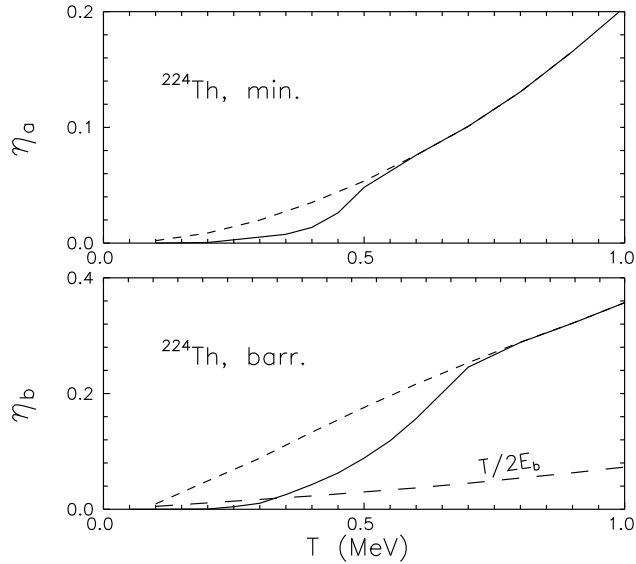}
}}
\caption{The damping factor $\eta=\gamma/(2\protect\sqrt{M\vert
C\vert})$ at the potential minimum (top) and at the fission
barrier (bottom) of $^{224}Th$; the dashed curves are for
$\Delta =0$. The long-dashed line in the bottom part shows the
function $T/2E_b$.}
\label{Fig.1}
\end{figure}

Let us demonstrate these features on the basis of numerical
calculations performed for the example of $^{224}{\rm Th}$. For
details we have to refer to \cite{ivhopa}, but we may mention that
the $\Gamma(\hbar\omega, \Delta, T)$ has been calculated along the
lines suggested in \cite{sijecop}. In Fig.1 we display the
$\eta(T)$'s at the minimum and the barrier. They have been
obtained for a $\Delta=\Delta(T)$ as determined by the gap
equation.  Unfortunately, so far it has not been possible to
calculate the underlying response function $\chi(\omega)$ in full
glory.  Rather, when evaluating the necessary folding integrals
over frequency the correct width had to be approximated by a
constant, calculated at the Fermi energy $\mu$:
$\Gamma(\hbar\omega, \Delta, T) = \Gamma(\hbar\omega = \mu,
\Delta, T)$. Indeed, in this regime of small $\omega$ and for
$T\leq T_{pair}$ where pair correlations disappear, such an
estimate may be considered to represent the correct width well
enough to allow for a general analysis. Evidently, the values of
$\eta$'s obtained for $\Delta\neq 0$ clearly fall well below those
of the unpaired case, shown here by the dashed lines.

The most important features exhibited in Fig.1 may be summarized
as follows, together with the consequences for Kramers' decay
rate: (1) The step like function mentioned before is clearly
visible. (2) Below $T\approx T_{pair} \simeq 0.5 {\;{\rm MeV}}$
the effective damping rate $\eta$ is smaller than about $0.1$. (3)
As seen in Fig.1, $\eta$ may fall below $T/2E_b$ such that formula
(\ref{krahivi}) no longer applies. (4) Up to $T\simeq 1{\;{\rm
MeV}}$ $\eta$ stays below $\simeq 0.2$ at the minimum and below
$\simeq 0.3$ at the barrier. The latter value implies that the
rate may be approximated fairly well by the transition state value
$r_{TST}\approx R_K(\eta=0)$, see (\ref{krahivi}). (5) These
values of $\eta$ are much smaller than those one gets within
"macroscopic models", say in terms of a combination of wall
friction with the stiffness and inertia of the liquid drop model
(with irrotational flow).

\begin{figure}[htb]
\centerline{{
\epsfxsize=8.5cm
\epsffile{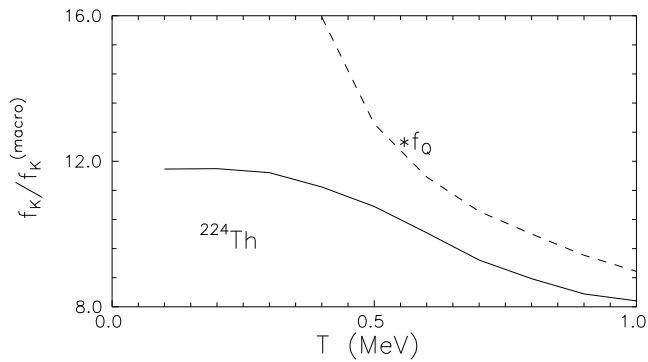}
}}
\caption{The ratio of the Kramers correction factor $f_K=$
$\protect\sqrt{1+\eta_b^2}-\eta_b$ to its macroscopic counterpart as
function of temperature. The quantity $f_Qf_K/f_k^{macro}$ accounts for
the quantum correction factor $f_Q$, see eq.(\protect\ref{quacofa}).
}
\label{Fig.2}
\end{figure}

In Fig.2 we plot the ratio of Kramers' pre-factor
$f_K=(\sqrt{1+\eta_b^2} - \eta_b) $ obtained from our results to
that of the macroscopic limit just described \cite{machigh}. It is
seen that the latter {\it underestimates} the decay rate by about
a factor of 10. Quantum corrections will increase this deviation
further, indicated here by the dashed curve.

\medskip

Within the LHA, these quantum corrections come in through the
diffusion coefficients, as given by the fluctuation dissipation
theorem \cite{hofrep}. It is only at temperatures above $2~{\;{\rm
MeV}}$ that one may safely assume the classic Einstein relation
$D_{pp}=\gamma T$ to be valid \cite{hofrep}. In the general case,
in addition to the $D_{pp}$ there is a cross term $D_{qp}$, both
of which depend in non-linear way on combinations of $M,~\gamma$
and $C$, or on the parameters introduced in (\ref{defgameta}). The
diffusion coefficients behave very differently for stable and for
unstable modes. To demonstrate this feature let us look at the
limit of small dissipation $\eta \ll 1$. To lowest order in
$\gamma$ one gets
\begin{equation}\label{diffus}
D_{qp} = 0~ D_{pp} = \gamma T^*~{\rm with }~
T^*(\varpi)={\hbar\varpi\over 2}\coth\left({\hbar\varpi\over 2T}\right)
\end{equation}
with $\varpi = \sqrt{C / M} =\vert\varpi\vert $ for $C>0$ and
$\varpi = i\vert\varpi \vert $ for $C<0$. The form (\ref{diffus})
may be said to represent the correct behavior fairly well below
$\eta\simeq 0.1$ (see Fig.3.4.2 of \cite{hofrep}). From the
results shown above for $\eta(T)$, one may thus argue the relation
(\ref{diffus}) to be acceptable for temperatures below $T_{pair}$,
whereas deviations must be expected for $T\geq T_{pair}$.  For
$C<0$ and weak friction the diffusion coefficient $D_{pp}$ falls
below the values given by the Einstein relation. It quickly drops
to zero at a critical temperature $T_c$, below  which the $D_{pp}$
would become negative and the diffusion equation would loose its
meaning. The value of this $T_c$ decreases with increasing $\eta$,
such that the form given in (\ref{diffus}) delivers an upper limit
and we may write
\begin{equation}\label{crittemp}
T_c\leq T_c(\eta=0) =  \hbar\varpi_b /\pi
  \qquad\qquad T_c < T_{pair}
\end{equation}
The statement on the right is reached assuming the $\hbar \varpi_b $
to be of the order of $1{\;{\rm MeV}}$ and taking the value for $T_{pair}$, as
reported
above, together with the fact that below  $T_{pair}$ the damping rate
$\eta$ falls below $0.1$.

Commonly, the quantum corrections to Kramers decay rate are
expressed by a factor $f_Q$ appearing in the correct rate as
$R=f_QR_K$ (see e.g. \cite{hantalbor}). As shown in
\cite{hofthoming}, this form may also be obtained within the LHA.
This derivation is based on the assumption that in the
neighborhood of the potential minimum friction is large enough to
ensure sufficient relaxation inside the well. The same assumption
is behind  Kramers' "high viscosity limit", upon which we just
have convinced ourselves to be given in the range of temperature
at $T_{pair}$ and above. Moreover, this assumption turns out to be
necessary also when the problem is formulated and solved with path
integrals in real time propagation (see \cite{retipropa}).

The $f_Q$ can be expressed by a ratio of two partition sums:
$ f_Q = \vert{\cal Z}_b\vert /{\cal Z}_a$, where the one associated to
the barrier has to be defined by analytic continuation.  According to
\cite{graweitalk}, the ${\cal Z}$ of a damped oscillator can be calculated
from the equilibrium fluctuations of momentum and coordinate. Hence,
within the LHA it might be expressed by the diffusion coefficients.
Unfortunately, for $\gamma \neq 0$ a calculation of the momentum
fluctuation requires regularization, for instance by introducing a
frequency dependent friction coefficient (Drude regularization).
To get a fairly simple estimate of $f_Q$ and its T-dependence
we used the following formula (with $\hbar\omega_n=n 2\pi T$)
\begin{equation}\label{quacofa}
f_Q = \prod_{n=1}^\infty {\omega_n^2 +
\omega_n\overline{\Gamma_\gamma} + \varpi_a^2 \over \omega_n^2 +
\omega_n\overline{\Gamma_\gamma} -\varpi_b^2} \quad {\rm with}
\quad \overline{\Gamma_\gamma}= { \Gamma_\gamma^a +
\Gamma_\gamma^b \over 2}
\end{equation}
It may be noted in passing, (a) that without (Drude)
regularization this formula would diverge for $\Gamma_\gamma^a
\neq \Gamma_\gamma^b$ and (b) that problems of this type are
absent for the Caldeira-Leggett approach where the transport
coefficients do not change with the collective variable;
generalizations are possible, though, for instance by introducing
variable coefficients phenomenologically, see e.g.
\cite{froebtil}. The result of a numerical evaluation of
(\ref{quacofa}) within our theory is shown in Fig.2 by the dashed
curve. This graph demonstrates several features, valid in this
range of temperatures: (i) The quantum effects in the collective
motion may change the decay rate by about 30 \% or less.  (ii)
Already at $T=1 {\;{\rm MeV}}$ they only amount to about 10~\%.
(iii) More important are the quantum effects of nucleonic motion,
which are responsible for the deviation of the transport
coefficients from the macroscopic models.

Unfortunately, it is not possible to carry the analysis further
down to smaller temperatures. Below $T_c$ [recall
(\ref{crittemp})] we have seen the LHA break down. It may be said
that this problem generally appears in real time formulations.
Within the functional integral approach this has been demonstrated
in \cite{retipropa} and traced back to the harmonic approximation
to the barrier.  The problem at stake here is a very severe one
for any applications of Langevin or Fokker Planck equations to
nuclear physics. Both methods allow one to account for various
non-linear effects, as manifested by variable transport
coefficients, for instance, but both of them rely on "real time
propagation". Moreover, practical computer programs exploit
locally harmonic approximations, in one way or other, such that it
is not possible to even define meaningful diffusion coefficients
below $T_c$.

In this context we should like to mention the possibility of
calculating and exploiting a Feynman-Vernon functional for global
motion on the basis of Random Matrix Theory \cite{buldoku}. In
principle, this might allow one to study quantum effects, but it
is somewhat questionable whether this procedure will be applicable
to nuclear physics, at those temperatures where these quantum
effects become important which we just mentioned. This concern has
essentially two reasons --- discarding for the moment the very
fact that this model, too, has difficulties with self-consistency.
(i) Generally an application of RMT ceases to be valid at low
excitations.  (ii) So far practical applications have been
possible only to leading order in an expansion in $1/T$, actually
to that regime of $T$ where friction decreases with $T$.

Summarizing our results we hope to have been able to exhibit for
low energy nuclear physics an exciting problem of quantum
transport which still is lacking a general solution. With respect
to the application in nuclear physics itself, it appears to be
very difficult to describe theoretically processes like the "cold"
production of super heavy elements without a decent understanding
of transport at small temperatures and weak dissipation. Whereas
low thermal excitations are dictated by experimental conditions,
the fact of small friction then is a consequence of the quantal
nature of nucleonic dynamics in a mean field, in particular when
pair correlations become important. Of course, to obtain more
quantitative results, further studies on the microscopic level are
needed. For instance, it is necessary to understand better the
mechanism of "collisions" under the presence of pair correlations.
Their role on the T-dependence of transport, as well as that of
the "heat pole" in the larger range of excitations, require
further clarification. Likewise, it should be very interesting to
allow for fluctuations in the gap parameter and to examine in
which way they might modify fission dynamics.

The authors like to acknowledge fruitful discussions with J.
Ankerhold and N.V. Antonenko. This work was supported by the
Deutsche Forschungsgemeinschaft.



\begin{thebibliography}{99}
\bibitem{temflu} To simplify matters we want to discard here the
problem of possible fluctuations of $T$, which
may be deduced from the formula $\Delta T/T= 1/\sqrt{{\cal C}}$.
Estimating the specific heat ${\cal C}$ from the Fermi gas
model one obtains $\Delta T/T = 1/\sqrt{AT}$ (T in MeV).
\bibitem{machigh} These macroscopic models are related to the high temperature limit
of our microscopic approach, see e.g. \cite{hofrep}.
\end{thebibliography}
\end{document}